# Fluid-particle interaction regimes during the evolution of turbidity currents from a coupled LES/DEM model

Jiafeng Xie[1], Peng Hu[2,*], Thomas Pähtz[3,*], Zhiguo He[4], Niansheng Cheng[5]

1 Research assistant, Ocean College, Zhejiang University, China. xiejf@zju.edu.cn

2 Corresponding author: Professor, Ocean College, Zhejiang University, China. pengphu@zju.edu.cn

3 Corresponding author: Research Professor, Ocean College, Zhejiang University, China. 0012136@zju.edu.cn

4 Professor, Ocean College, Zhejiang University, China. hezhiguo@zju.edu.cn

5 Professor, Ocean College, Zhejiang University, China. nscheng@zju.edu.cn

*corresponding authors

**Abstract:** In this paper, fluid-particle interactions in lock-exchange turbidity currents (TCs) over a flat bed are investigated using a model combining LES and DEM (Large-Eddy Simulation and Discrete Element Method). The reliability of this model is demonstrated via comparing the numerical solutions with measurements of the front positions, fluid velocity profile, and particle concentration profile of lock-exchange TCs. The following physical understandings are obtained. The vorticity field plays an important role for the current evolution by affecting the fluid lift force (i.e., in the direction normal to the fluid-particle slip velocity) acting on the particles. At the very beginning, a longitudinal positive lift force due to strong positive vorticity promotes longitudinal particle transport. Afterwards, the longitudinal lift force decreases and eventually becomes negative, with a magnitude that even exceeds that of the positive longitudinal drag force, because more and more of the settling particles are affected by the negative vorticity near the bottom wall caused by surface friction. Interestingly, in spite of the complex behavior of the fluid-particle interaction forces and their role in TC evolution, only a very small fraction of the initial particle gravitational potential energy is actually transformed into TC kinetic energy (both particle and fluid).

# 1 Introduction

Gravity flows are ubiquitous in nature, such as rock and snow avalanches, debris flows, turbidity currents and plumes in the ocean and reservoirs, and sea breezes in the atmosphere (Simpson, 1982; Qian and Wan, 1983). A turbidity current (TC) is a special kind of gravity flow whose driving force comes from longitudinal pressure gradients caused by spatial variations of the suspended particle load (Meiburg et al., 2015). As a common phenomenon in the deep ocean floors and river reservoirs, it has significant impacts on the exploration of oil (Meiburg and Kneller, 2010; Weimer et al., 2007) and gas and the stability of submarine structures (Zakeri et al., 2008; Meiburg et al., 2015; Kneller et al., 2016), as well as reservoir sedimentation (Wright and Marineau, 2019; Hsu et al., 2017).

Mathematical models of TCs can be divided into two categories: layer-averaged models and depth-resolving models (Meiburg et al., 2015). The former include box models (e.g., Dade and Huppert, 1995; Gladstone and Woods, 2000) and shallow water models (e.g., Parker et al., 1986; Parker et al., 1987; Bonnecaze et al., 1993; Hu and Cao, 2009; Hu et al., 2012; Hu et al., 2015; Liu et al., 2017; Wang et al., 2017; Yang et al., 2019; Hu et al., 2019; Alhaddad et al., 2020). The latter include Reynolds-averaged Navier-Stokes (RANS) equations (e.g., Huang et al., 2007, 2008; Abd El-Gawad et al., 2012; Felix, 2001; An and Li, 2010), Large Eddy Simulations (LES) (e.g., An et al., 2012; Kyrousi et al., 2018), and Direct Numerical Simulations (DNS) (e.g., Necker et al., 2002, 2005; Strauss and Glinsky, 2012; Espath et al., 2014; Nasr-Azadani and Meiburg, 2014; Francisco et al., 2017; He et al., 2018; He et al., 2019; Zhao et al., 2019). Depth-resolving models can be further subdivided into quasi-single-phase and two-phase models. In quasi-single-phase models, the particle phase is represented by a continuous particle concentration field and assumed to exhibit the same velocity as the fluid phase, or a velocity mapped to the fluid phase velocity using an empirical relationship (e.g., Decrop et al., 2015), while the mass and momentum balances of the particle-fluid mixture control the TC evolution. In two-phase models, particle and fluid phase are solved separately from one another, with separate mass and momentum conversation equations for each phase. In the past, most two-phase models of TCs have been Eulerian-Eulerian models (e.g., Meiburg et al., 2015; Lee, 2019). However, while these models have made significant contributions to the understanding of turbidity currents (Meiburg et al., 2015), one can achieve a higher degree of physical realism by resolving individual particles. In particular, only a material description of the particle phase can accurately characterize the transport and deposition behavior of individual suspended particles and thus fluid-particle interactions.

This study aims to characterize the flow processes, fluid-particle interactions, and energy conversion processes that are caused by the complex deposition dynamics of suspended particles during the evolution of a TC. The study employs a coupled Large Eddy Simulation - Discrete Element Method (LES-DEM) model, provided and coded by previous studies (e.g., Kloss et al., 2012). The model belongs to CFD-DEM models, which have had considerably improved our understanding of particle-laden flows in recent years (e.g., Tsuji et al., 1993; Peng et al., 2014; Tsuji et al., 2008; Deen et al., 2007; Smuts, 2015; Deshpande et al., 2019; Zhu et al., 2007, 2008; Jing et al., 2016; Jing et al., 2018; Zhang and Yin, 2018), most notably sediment transport (Zhao and Shan, 2013; Li and Tao, 2018; Li and Zhao, 2018; Sun et al., 2018; Sun and Xiao, 2016a, 2016b; Furbish and Schmeeckle, 2013; Schmeeckle, 2014; Zhao et al., 2014; Lu et al., 2019; Zhang et al., 2020; Durán et al., 2012; Maurin et al., 2015; Maurin et al., 2018; Pähtz and Durán, 2018a, 2018b, 2020), though their application to TCs have been rather limited (Meiburg et al., 2015; Biegert et al., 2017). In CFD-DEM models, a Navier-Stokes equations-based description of hydrodynamics is coupled with the DEM for particles (Sun and Sakai, 2015).

The reminder of the paper is organized as follows. The introduction and implementation of the model are presented in Section 2. The mathematical model setting and simulation results are given in Section 3, including the validation of the simulated TC flow field, and particle characteristics. Section 4 discusses the interphase relationship from the formation and evolution of TCs, the head dynamics, and the energy conversion process. Section 5 is the summary of this study.

**2 Methodology**

2.1 Governing equations for the fluid phase

The incompressible Navier-Stokes equations are adopted for the fluid phase, including the continuity equation and momentum conservation equation (e.g., Zhang and Yin, 2018; Chu et al., 2009):

$$\frac{\partial(\alpha_f \rho_f \mathbf{u}_f)}{\partial t} + \nabla \cdot (\alpha_f \rho_f \mathbf{u}_f \mathbf{u}_f) = -\nabla p + \alpha_f \rho_f \mathbf{g} + \alpha_f \nabla \cdot \mathbf{\tau} - \mathbf{R}_{pf} + \nabla \cdot (\alpha_f \mathbf{\Gamma}) \quad (1a)$$

$$\frac{\partial \alpha_f}{\partial t} + \nabla \cdot (\alpha_f \mathbf{u}_f) = 0 \quad (1b)$$

where $α_f$ is the fluid volume fraction ($α_f = 1 - α_p$), with $α_p$ is the particle volume fraction, $ρ_f$ is the fluid density, $\mathbf{u}_f$ is the fluid velocity, $p$ is the fluid pressure, $\mathbf{g}$ is the gravitational acceleration, $\boldsymbol{τ}$ is the viscous stress tensor of the fluid, $\mathbf{R}_{pf}$ is the fluid-particle interaction force per unit volume, which is obtained by solving the governing equations for the particle phase, and $\boldsymbol{Γ}$ is the sub-grid stress tensor of fluid phase, which is resolved by the Smagorinsky model (SGS model). Note that the fluid velocity, fluid pressure, fluid stress tensor, and fluid sub-grid stress tensor are filtered variables.

2.2 Governing equations for particles

For the particle phase, the translational and rotational motions of each particle are controlled by Newton's second law and the conservation of angular-momentum (e.g., Gui et al., 2018; Chassagne et al., 2020):

$$m_i \frac{d\mathbf{u}_{p,i}}{dt} = \sum_{j=1}^{n_i^c} \mathbf{F}_{ij}^c + \mathbf{F}_i^f + m_i \mathbf{g} \tag{2a}$$

$$I_i \frac{d\boldsymbol{ω}_{p,i}}{dt} = \sum_{j=1}^{n_i^c} \mathbf{M}_{ij}^c \tag{2b}$$

where $m_i$ is the mass of the particle $i$, $\mathbf{u}_{p,i}$ is the translational velocity of the particle $i$, $\mathbf{F}_{ij}^c$ is the contact force acting on particle $i$ by particle $j$ or the boundary wall, $\mathbf{F}_i^f$ is the fluid-particle interaction force acting on particle $i$, $\boldsymbol{ω}_{p,i}$ is the angular velocity of particle $i$, $I_i$ is the moment of inertia of particle $i$, $\mathbf{M}_{ij}^c$ is the torque acting on particle $i$ by particle $j$ or the boundary wall, and $n_i^c$ is the number of particles contacting particle $i$. In order to determine the contact force $\mathbf{F}_{ij}^c$, the particles are modeled as soft spheres using an elastic spring and a viscous dashpot (Cundall and Strack, 1979). The contact force is calculated as follows:

$$\mathbf{F}_{ij}^c = \mathbf{F}_{ij}^n + \mathbf{F}_{ij}^t \tag{3}$$

where $\mathbf{F}_{ij}^n$ and $\mathbf{F}_{ij}^t$ are the normal component and the tangential component of the contact force, respectively, which can be expressed as:

$$\mathbf{F}_{ij}^n = k_n δ_{nij} - γ_n v_{nij} \tag{4a}$$

$$\mathbf{F}_{ij}^t = k_t δ_{tij} - γ_t v_{tij} \tag{4b}$$

where $k_n$ and $k_t$ are the elastic constants for normal contact and tangential contact, respectively, $δ_{nij}$ and $δ_{tij}$ are the normal overlap distance and tangential displacement vector between particle $i$ and particle *j*, respectively, $\gamma_n$ and $\gamma_t$ are the viscoelastic damping constants for normal contact and tangential contact, respectively, and $v_{nij}$ and $v_{tij}$ are the normal component and tangential component of the relative velocity of particle *i* and particle *j*, respectively. The values of the elastic and damping constants chosen in this paper correspond to a Young modulus of $5\times10^6$ Pa, Poisson ratio $v = 0.45$ and normal and tangential restitution coefficients of $e = 0.3$. When $\mathbf{F}_{ij}{}^t < \mu\mathbf{F}_{ij}{}^n$, where $\mu = 0.6$ is the friction coefficient, the tangential contact force $\mathbf{F}_{ij}{}^t$ is calculated by Eq. (4b); otherwise, it is equal to $\mu\mathbf{F}_{ij}{}^n$.

2.3 Fluid-particle interactions

The momentum change $\mathbf{R}_{pf}$ in Eq. (1a) and the fluid-particle interaction force $\mathbf{F}_i{}^f$ for particle *i* in Eq. (2a) are related to each other via

$$\mathbf{R}_{pf} = \frac{\sum_{i=1}^{k_c} \mathbf{F}_i^f}{V_c} \tag{5}$$

where $V_c$ is the volume of a CFD cell, and $k_c$ is the number of particles contained in it. Since this work mainly focuses on the predominant physical processes, rather than developing a fully predictive model, and since existing literature expressions for particle-fluid interaction forces are largely empirical, we incorporate only the predominant fluid-particle interactions in $\mathbf{F}^f$: buoyancy ($\mathbf{F}^b$), drag ($\mathbf{F}^d$), lift ($\mathbf{F}^l$), and added mass ($\mathbf{F}^{add}$). Other forces are typically more than an order of magnitude smaller, and often difficult to describe analytically - especially the Basset force - (e.g., see the discussions in Section 2.3 of Schmeeckle (2014), Appendix A of Durán et al. (2011), and Appendix D of Pähtz et al. (2021)).

For non-sloped systems, for which the pressure can be approximated by the hydrostatic pressure, the buoyancy acceleration is approximately aligned with the gravity acceleration (i.e., $\mathbf{g} = (0, 0, -9.81)$ m/s$^2$), and the buoyancy force $\mathbf{F}^b$ is therefore approximately given by:

$$\mathbf{F}^b = -\frac{1}{6}\pi\rho_f d_p{}^3 \mathbf{g} \tag{6}$$

where $d_p$ is the particle diameter.

The drag force is estimated as (Di Felice, 1994):

$$\mathbf{F}^d = \frac{1}{8}C_D\rho_f \pi d_p{}^2 (\mathbf{u}_f - \mathbf{u}_p)|\mathbf{u}_f - \mathbf{u}_p|\alpha_f{}^{1-\chi} \tag{7}$$

where $C_D$ is the drag coefficient, obtained as:

$$C_D = \left(0.63 + \frac{4.8}{\sqrt{\alpha_f \text{Re}_p}}\right)^2 \quad (8)$$

where the particle Reynolds number $\text{Re}_p$ is given by:

$$\text{Re}_p = \frac{\rho_f d_p |\mathbf{u}_f - \mathbf{u}_p|}{\mu_f} \quad (9)$$

The parameter χ in Eq. (7) is a correction factor, which depends on $\text{Re}_p$ (Di Felice, 1994):

$$\chi = 3.7 - 0.65 \exp\left[-\frac{(1.5 - \log_{10}(\alpha_f \text{Re}_p))^2}{2}\right] \quad (10)$$

for $0.4 \leq \alpha_f \leq 1$ and $10^{-4} \leq \text{Re}_p \leq 10^6$.

The lift force model is calculated as (McLaughlin, 1991; Loth and Dorgan, 2009):

$$\mathbf{F}^l = \frac{1}{8}\pi \rho_f d_p^2 C_L |\mathbf{u}_f - \mathbf{u}_p| \left[(\mathbf{u}_f - \mathbf{u}_p) \times \frac{\boldsymbol{\omega}_f}{|\boldsymbol{\omega}_f|}\right] \quad (11)$$

where $\boldsymbol{\omega}_f$ is the fluid vorticity, and $C_L$ is the lift coefficient given by (McLaughlin, 1991; Loth and Dorgan, 2009):

$$C_L = J^* \frac{12.92}{\pi}\sqrt{\frac{\omega^*}{\text{Re}_p}} + \Omega^*_{p,eq} C^*_{L,\Omega} \quad (12)$$

with

$$\omega^* = \frac{|\boldsymbol{\omega}_f| d_p}{|\mathbf{u}_f - \mathbf{u}_p|} \quad (13)$$

The function $J^*$ in Eq. (12) reads (Mei, 1992)

$$J^* = 0.3\left\{1 + \tanh\left[\frac{5}{2}\left(\log_{10}\sqrt{\frac{\omega^*}{\text{Re}_p}} + 0.191\right)\right]\right\}\left\{\frac{2}{3} + \tanh\left[6\sqrt{\frac{\omega^*}{\text{Re}_p}} - 1.92\right]\right\} \quad (14)$$

Furthermore, the empirical correction $C^*_{L,\Omega}$ and empirical model for $\Omega^*_{p,eq}$ in Eq. (12) are given by (Loth and Dorgan, 2009):

$$\Omega^*_{p,eq} = \frac{\omega^*}{2}(1 - 0.0075\text{Re}_\omega)(1 - 0.062\sqrt{\text{Re}_p} - 0.001\text{Re}_p) \quad (15)$$

$$C^*_{L,\Omega} = 1 - \{0.675 + 0.15(1 + \tanh[0.28(\Omega^*_p - 2)])\}\tanh[0.18\sqrt{\text{Re}_p}] \quad (16)$$

where $\text{Re}_\omega$ and $\Omega^*_p$ are expressed, respectively, as:

$$\text{Re}_\omega = \frac{\rho_f |\boldsymbol{\omega}_f| d_p^2}{\mu_f} \tag{17}$$

$$\Omega^*_p = \frac{|\boldsymbol{\omega}_p| d_p}{|\mathbf{u}_f - \mathbf{u}_p|} \tag{18}$$

Note that this lift force formulation requires $\text{Re}_p < 50$ to be applicable (Loth, 2008; Loth and Dorgan, 2009), which we confirmed is much larger than the range of $\text{Re}_p$ in our TC simulations (0.01-0.1). Furthermore, note that the lift force formulation accounts for vorticity-induced lift (shear lift) and lift due to particle rotation (Magnus lift).

The added mass force is modeled as (e.g., Sun and Xiao, 2016b; Zhang et al., 2021):

$$\mathbf{F}^{add} = \frac{1}{6} C_{add} \rho_f \pi d_p^2 \left( \frac{D\mathbf{u}_f}{Dt} - \frac{D\mathbf{u}_p}{Dt} \right) \tag{19}$$

where $C_{add} = 0.5$ is the added mass coefficient, $D\mathbf{u}_f/Dt = \partial \mathbf{u}_f/\partial t + \mathbf{u}_f \cdot \nabla \mathbf{u}_f$ the material derivative of the fluid velocity interpolated to the position of particle center, and $D\mathbf{u}_p/Dt$ the particle acceleration.

Fluid-induced torque is not considered, since the grid cells, over which variables are averaged, are much larger than the particle spacing (see Section 2.3 of Jing et al. (2016) for details).

**2.4 Model implementations**

In the present LES-DEM model, the fluid module is based on the open source code OpenFOAM (OpenCFD, 2016), and the particle module adopts the open source particle-simulating software LIGGGHTS, which was initially developed by Sandia National Laboratories for molecular dynamics. The coupling between particles and fluid is based on the open source CFD-DEM engine CFDEMproject, documented and completely coded by Kloss et al. (2012) (also see https://www.cfdem.com).

The fluid module (as given in Section 2.1) is solved using the finite volume method. Local average physical quantities, such as pressure, velocity, and fluid volume fraction, are stored at the centers of the grid cells. The contribution of a particle within a given grid cell to the cell average is weighted by the fraction of this particle's volume that is contained within this cell. Furthermore, linear interpolation is used to interpolate the data from the grid center to face center. This averaging-interpolation scheme causes a smoother behavior of the average fields when compared with other common schemes.

For Equations (1a) and (1b), the temporal term is discretized using the first-order implicit Euler scheme, and the spatial discretization of the gradient term, divergence term, and Laplacian term are all based on second-order linear interpolation. The computational mesh size is generally larger than the diameter of the particles.

**3 Simulation of a turbidity current**

This section is separated into several subsections. Section 3.1 describes the initial numerical setup that we use to simulate the TC. Section 3.2 evaluates the evolution of two global TC variables of the simulated TC, the front position and velocity, and validates the simulation with experimental data. We then turn our attentions to the fluid flow in Section 3.3 and the particle flow in Section 3.4, the understanding of both of which is necessary for the evaluation of the role of fluid-particle interactions during the evolution of the TC (Section 4).

3.1 Numerical setup

As shown in Figure 1, a cuboid numerical tank is considered: the length $L_x$, width $L_y$, and height $L_z$ are 75 mm, 5 mm, and 10 mm (i.e., $1500d_p$, $100d_p$, and $200d_p$), respectively. A fictitious gate is put $x_0 = 10$ mm (i.e., $200d_p$) from the left-end wall. Initially, the tank is filled with water ($\rho_f = 1000$ kg/m$^3$), and particles ($\rho_p = 1200$ kg/m$^3$, $d_p = 0.05$ mm) are uniformly placed within the left reservoir. The initial particle concentration $C_0$ is 0.01 m$^3$/m$^3$, corresponding to a total particle number of $N_p = 76394$. When the gate is lifted, the fluid containing particles contacts the ambient fluid and moves along the bottom of the tank. The longitudinal ($x$) and transverse ($y$) walls exhibit no-slip boundary conditions and periodic boundary conditions, respectively, while for the bottom wall and the top wall ($z$ direction, parallel to gravity), a no-slip and free-slip boundary condition, respectively, are employed. The mesh sizes in $x$, $y$, and $z$ directions are $3d_p$, $4d_p$, and $2d_p$, respectively. The simulation duration is 5.0 s.

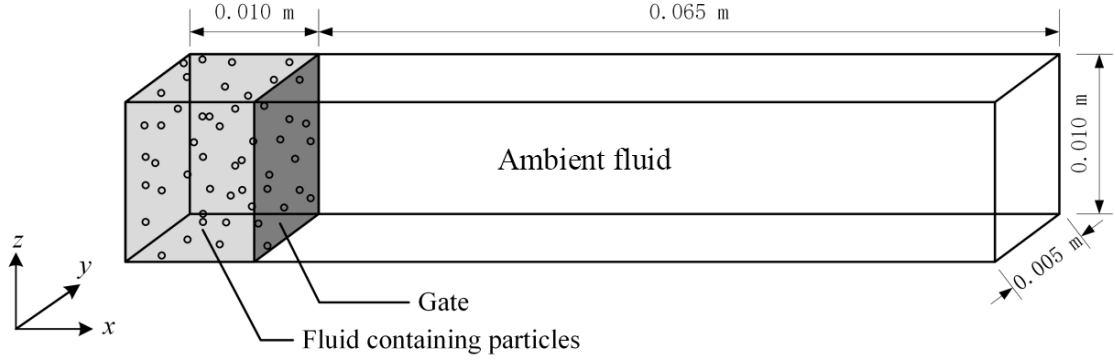

Figure 1. Schematic diagram for LES-DEM modeling of the TC.

3.2 Verification of TC numerical model

The front position $x_{front}$ of the current is defined by a threshold particle volume fraction $α_p = 10^{-5}$, corresponding to a situation where only 0.05% of the volume of a single particle is contained within a grid cell. Figures 2(a)-2(d) show the time evolutions of $x_{front}$ (consistent with experimental data by Gladstone et al. (1998)), the time evolution of the front velocity $u_{front}$, the dimensionless fluid longitudinal velocity profile at the time instant $t = 5.0$ s, and the dimensionless particle concentration at $t = 5.0$ s, respectively. Note that the exact choice of the time instant, provided it is near the end of the simulation, does not significantly affect the comparison with experiments below. Parameters in Figures 2(a)-2(b) are normalized using half of the water depth $L_z / 2$ as the characteristic length and the buoyancy velocity $u_b$ as the characteristic velocity (Francisco et al. 2017), where $u_b$ is given as:

$$u_b = \sqrt{g\frac{(\rho_p - \rho_f)C_0}{\rho_f}\frac{L_z}{2}} \qquad (20)$$

It can be seen in Figure 2(b) that the simulated TC process can be divided into two phases. During the initial generation phase (i.e., the acceleration phase at 0~0.7 s), the front velocity increases rapidly. Then, in the slumping phase (at 0.7~5.0 s), the front velocity is roughly constant (decreases slightly).

The layer-averaged height of the TC $H$ in Figures 2(c)-2(d) is defined as (Ellison and Turner, 1959):

$$H = \frac{\left(\int_0^\infty u_f(z)dz\right)^2}{\int_0^\infty u_f(z)^2 dz} \tag{21}$$

Note that the upper limit (infinity) of the integral in Eq. (21) should be interpreted as the elevation at which the fluid longitudinal velocity becomes zero (Imran et al., 2004). Moreover, $u_f^{\max}$ in Figures 2(c)-2(d) denotes the maximum fluid longitudinal velocity, $H_m$ is its elevation, and $\alpha_p^m$ is the particle concentration at $z = H_m$. The second-order central moment *SCM* of the experimental or simulated data in Figures 2(c)-2(d) are shown in Table 1, which is defined by:

$$SCM = \frac{1}{n_{sam}} \sum_{i=1}^{n_{sam}} \left(\tilde{A}_i - \overline{\tilde{A}}\right)^2 \tag{22}$$

where $n_{sam}$ is the number of samples, $\tilde{A}_i$ is the dimensionless data (dimensionless fluid velocity or dimensionless particle concentration), and $\overline{\tilde{A}}$ is the average of dimensionless data. It can be seen from Figure 2(c) that the fluid longitudinal velocity profile is consistent with experimental data (Hitomi et al., 2021; Farizan et al., 2019; Nourmohammadi et al., 2011; Altinakar et al., 1996), and *SCM* of the fluid longitudinal velocity data in present study is approximately the same as that in previous studies (Table 1). In Figure 2(d), the particle concentration profile is qualitatively consistent with experimental data (Hitomi et al., 2021; Farizan et al., 2019; Altinakar et al., 1996), but slight quantitative deviations occur. The present second-order central moments of particle concentration data are within the range of those in previous studies (Table 1).

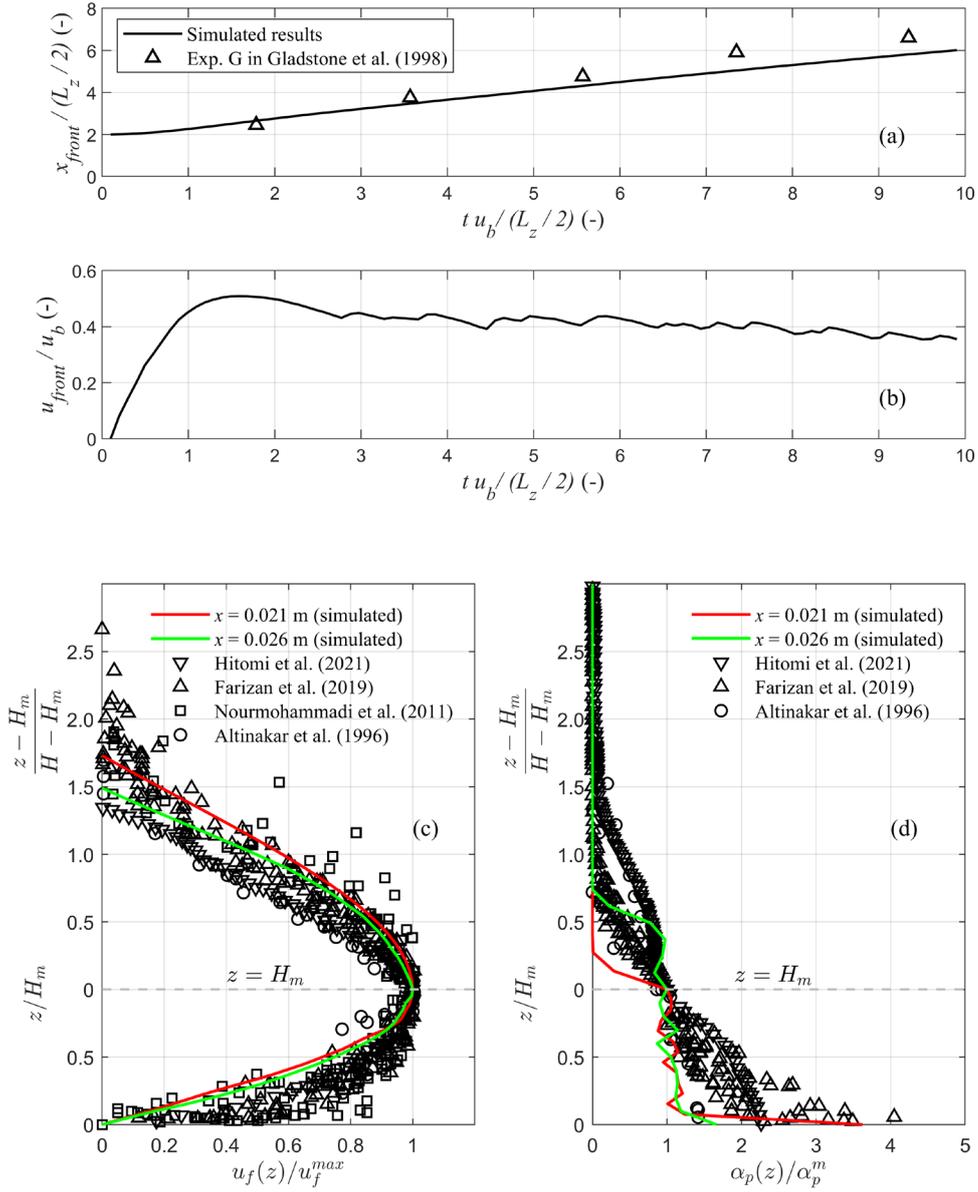

Figure 2. Temporal variations of (a) the dimensionless front position of the TC and (b) the dimensionless velocity of the TC front. Profiles of (c) the dimensionless fluid longitudinal velocity and (d) the dimensionless particle concentration at $t$ = 5.0 s.

Table 1. Second-order central moment *SCM* of data that are presented in Figures 2(c)-2(d).

|  | *SCM* | |
| --- | --- | --- |
|  | Fluid longitudinal velocity data | Particle concentration data |
| Hitomi et al. (2021) | 0.094 | 0.404 |

| | | | |
|---|---|---|---|
| Farizan et al. (2019) | | 0.104 | 0.761 |
| Nourmohammadi et al. (2011) | | 0.079 | - |
| Altinakar et al. (1996) | | 0.090 | 0.215 |
| In present study | $x$ = 0.021 m | 0.102 | 0.613 |
| | $x$ = 0.026 m | 0.094 | 0.230 |

3.3 Characteristics of the flow field

    Figure 3 shows the spatial distributions of the fluid longitudinal velocity in the *x-z* plane along with the fluid velocity vectors at six selected times. For convenience of description, the computational domain is divided into two regions: the ambient flow region (i.e., negligible sediment presence) (Area I) and TC region (Area II). The boundary between these two regions (i.e., the upper interface of the TC) is defined using the threshold concentration ($\alpha_p$ = 10$^{-5}$). For the TC region, the position of the maximum positive longitudinal velocity is near the front of the TC. As the current evolves, the maximum positive longitudinal velocity first increases (Figures 3(a)-3(c)) and then decreases (Figures 3(c)-3(f)). In the ambient flow region, the lower fluid layers are pushed by the front of the TC, except very near the bottom wall where friction resists this push. Hence, the TC develops a tongue-like shape. In particular, as the higher-density TC penetrates the lower-density ambient flow in the lower layers, the reverse process occurs in upper layers due to mass conservation. This results in a counterclockwise loop formed by the velocity streamlines (Figure 3). The size of this loop gradually increases with time. Note that the present simulations do not produce inflectional instabilities akin to the inviscid Kelvin-Helmholtz (K-H) mechanism. It is because the Richardson number $Ri = g'H/U^2$ in the present study is in the order of $O(10)$, in which $U$ is the mean fluid longitudinal velocity of the TC and $g' = (\rho_m - \rho_f)|\mathbf{g}|/\rho_f$ the effective gravitational acceleration ($\rho_m$ is the density of the particle-water mixture), whereas instabilities in turbidity currents often require that the Richardson number falls below a critical value of order unity (Turner, 1986). Nevertheless, we confirmed that a current process with larger particle concentration and/or particle density will result in a lower $Ri$ and the generation of inflectional instabilities.

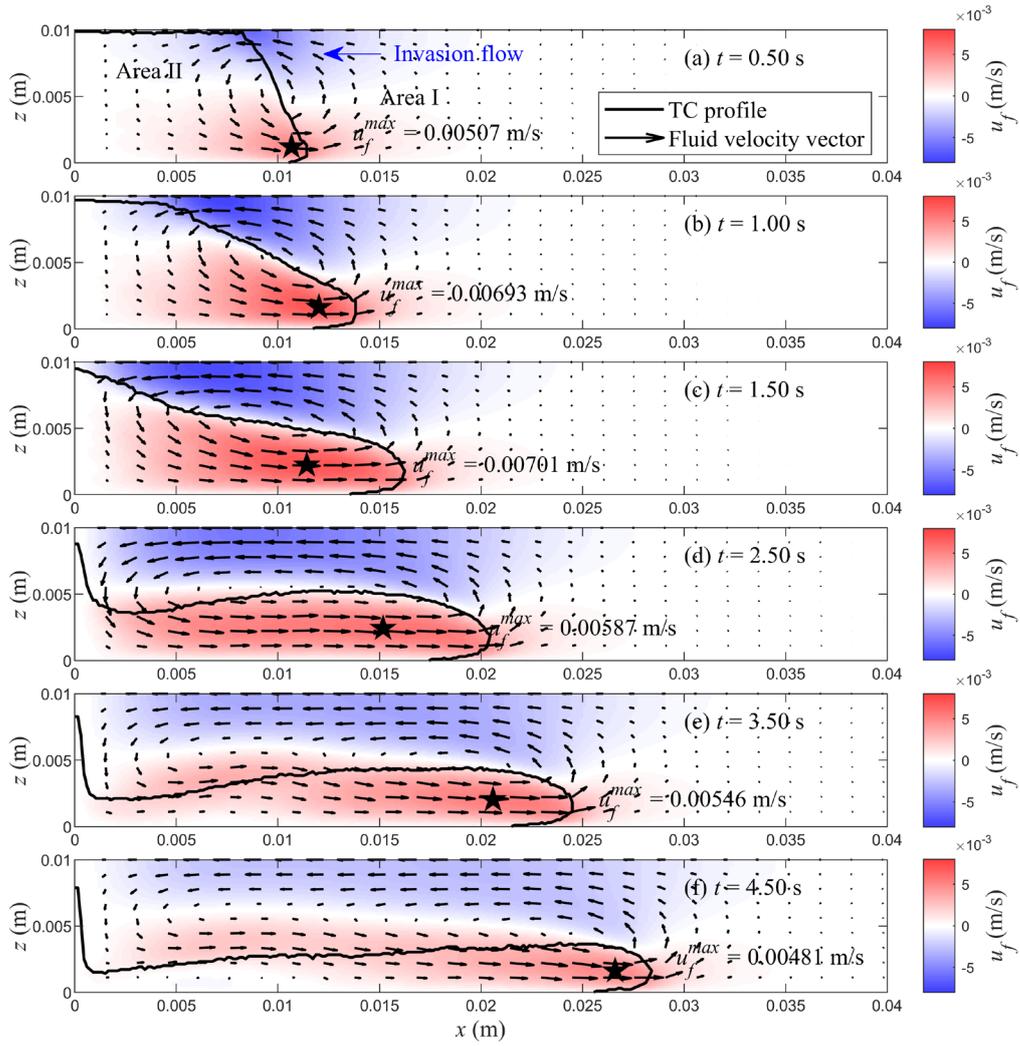

Figure 3. Fluid longitudinal velocity fields in the *x*-*z* plane at six selected times. The black arrows represent the fluid velocity vectors. The black stars indicate the location of the maximum fluid positive longitudinal velocity. The solid line indicates the TC profile, defined by a particle concentration threshold of $\alpha_p = 10^{-5}$.

Figure 4 shows the corresponding vorticity field in the *x*-*z* plane at the same selected times. At all times, positive vorticity, caused by the overall counterclockwise fluid motion, dominates the high fluid layers, whereas negative vorticity, caused by friction with the bottom wall, dominates the lower fluid layers. Both the maximum positive and maximum negative vorticity are located near the TC boundary. Note that the behavior of the vorticity is key to understand the behavior of the lift force acting on the particles (see Eq. (11)).

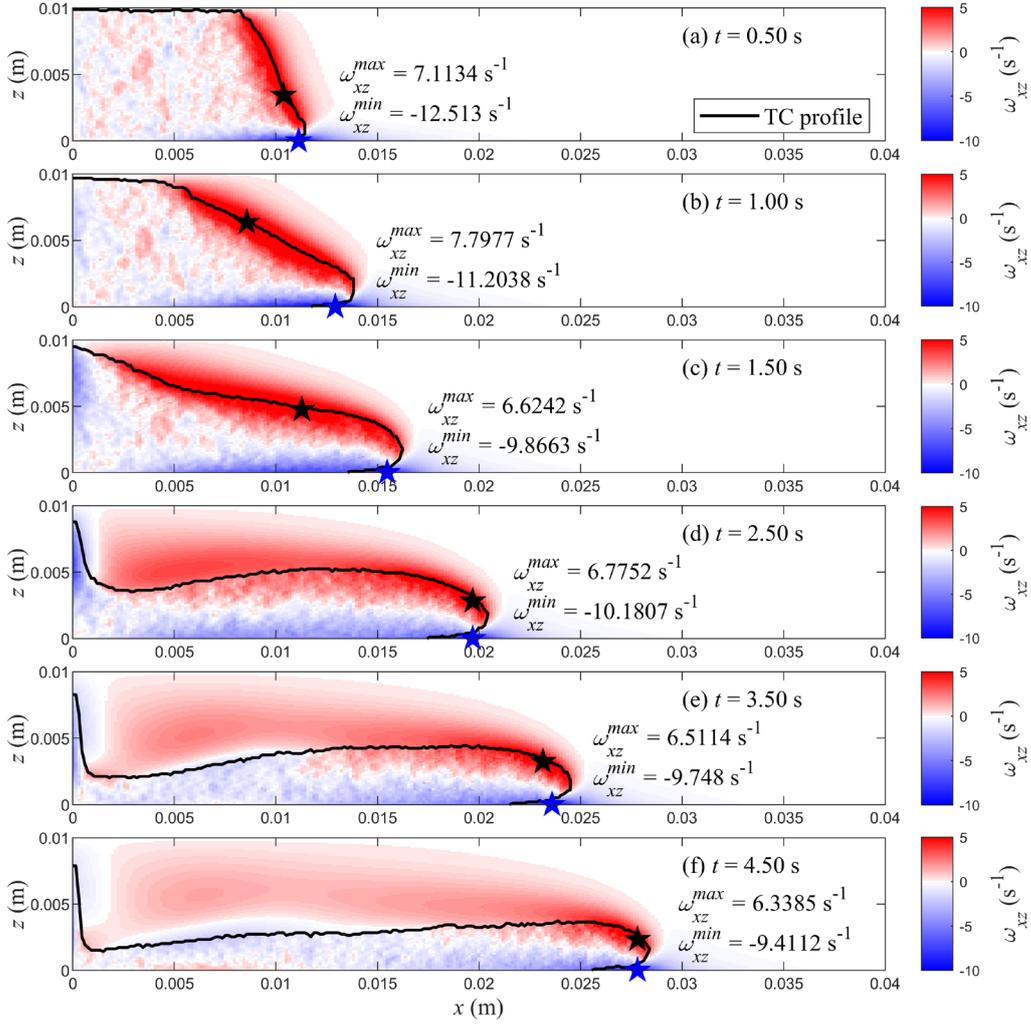

Figure 4. Fluid vorticity fields in the *x-z* plane at six selected times. $\omega_{xz} = \partial u_f / \partial z - \partial w_f / \partial x$. The black and blue stars indicate the location of the maximum vorticity and minimum vorticity, respectively, and the values are also indicated on the diagram. The solid line indicates the TC profile, defined by a particle concentration threshold of $\alpha_p = 10^{-5}$.

The *Q*-criterion can help to study the effect of the particle motion on the flow, such as investigating the lobe-and-cleft patterns (Espath et al., 2014; Francisco et al., 2017). Here, we have used this criterion to study coherent structures that may emerge during the evolution of the TC. Coherent structures reflect the flow topology of the TC, which can characterize the enstrophy-dominated flow structure when *Q* > 0.

The second invariant of the velocity gradient tensor *Q* includes the rotation rate tensor and strain rate tensor and is defined as (da Silva and Pereira, 2008):

$$Q = -\frac{1}{2}\left[\left(\frac{\partial u_f}{\partial x}\right)^2 + \left(\frac{\partial v_f}{\partial y}\right)^2 + \left(\frac{\partial w_f}{\partial z}\right)^2\right] - \frac{\partial u_f}{\partial y}\frac{\partial v_f}{\partial x} - \frac{\partial u_f}{\partial z}\frac{\partial w_f}{\partial x} - \frac{\partial v_f}{\partial z}\frac{\partial w_f}{\partial y} \quad (23)$$

Figure 5 shows coherent structures, obtained using a threshold value of $Q = 0.5$ s$^{-2}$ (equal to approximately 0.05 times the maximum $Q$ value at time zero), at six selected times (the reason for $Q = 0.5$ s$^{-2}$ is discussed in Appendix A). As shown in Figure 5, outside the TC region, at the beginning, there is a structurally complete and large-sized coherent structure (Figures 5(a)-5(c)), which becomes separated into two complete structures following the time evolution of the current (Figures 5(d)-5(e)). The upstream structure gradually decreases in size and eventually disappears (Figure 5(f)). In contrast, the downstream structure near the head, though it decreases in size to a certain extent, remains intact until the end of the simulation. Inside the TC region, especially near the TC interface (including upper and lower interfaces), there are many relatively small coherent structures, caused by the motion of individual particles. Note that as the TC evolves, the coherent structures near the interface between the ambient region and the TC region will gradually evolve into many finely fragmented structures, which has also been reported in previous studies (e.g., Pelmard et al., 2020; Ooi et al., 2007).

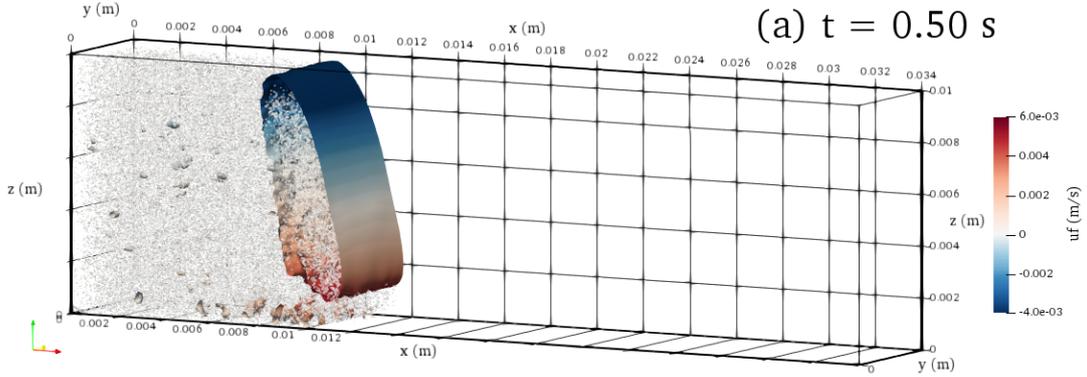

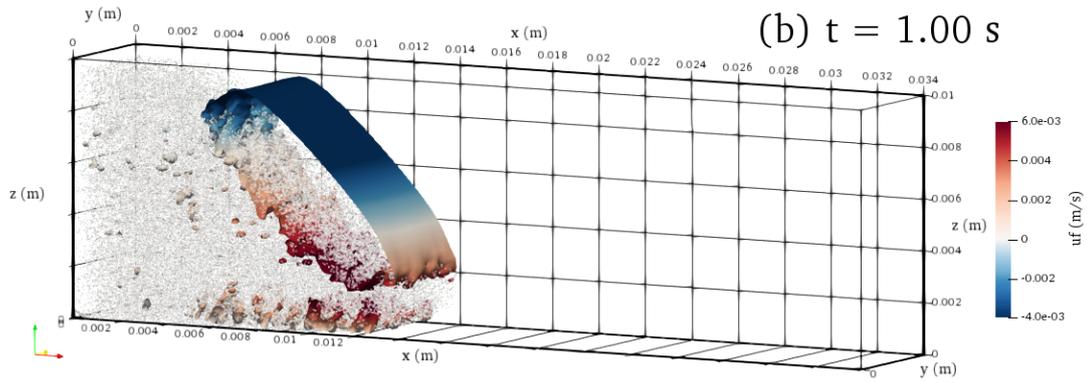

(b) t = 1.00 s

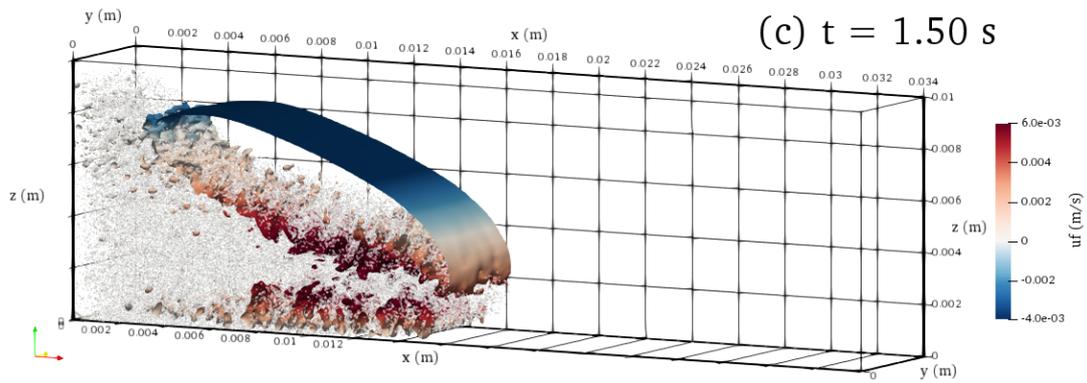

(c) t = 1.50 s

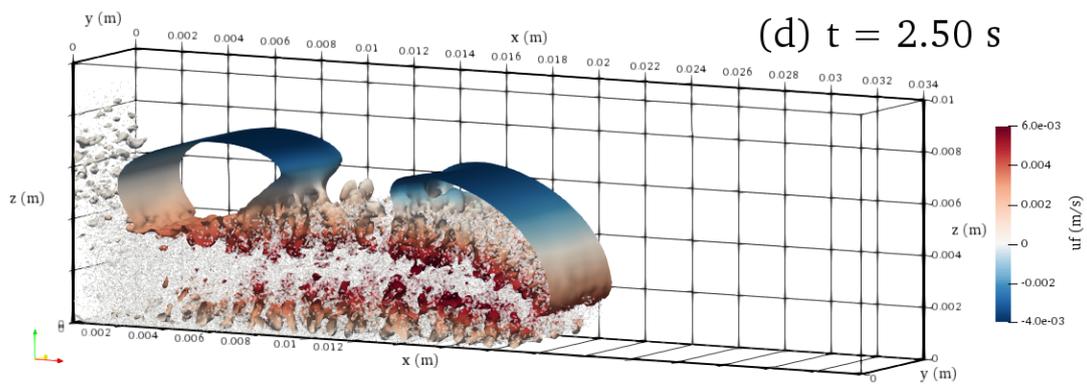

(d) t = 2.50 s

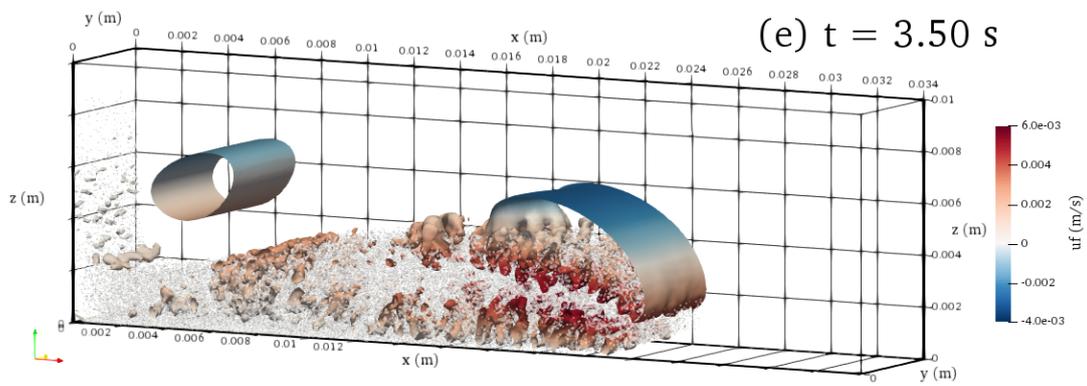

(e) t = 3.50 s

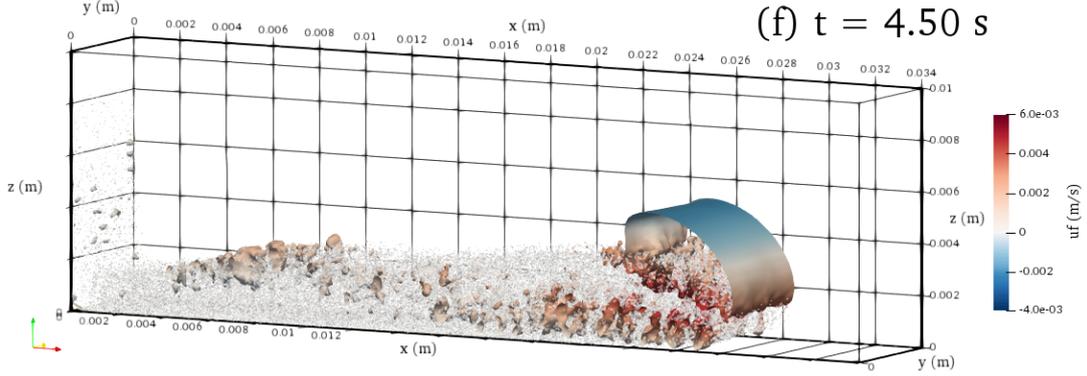

Figure 5. The coherent structures obtained by using the *Q*-criterion at six selected times. The iso-surface in the figure corresponds to $Q = 0.5$ s$^{-2}$, and the color of the iso-surface represents the longitudinal velocity of the fluid. The very small grey dots indicate particles.

3.4 Characteristics of particle transport and deposition

We decompose the particle vertical velocity $w_p$ into two terms: the terminal settling velocity $\omega_p$, calculated from Eq. (7) for a stationary flow at rest and the remaining particle vertical velocity $w_p - \omega_p$. The comparison between the two velocities can reflect whether the vertical movement of the particles is dominated by the fluid force in a stationary flow (background settling regime) or whether a more complex flow-particle interaction dynamics is at play (complex regime). When the absolute ratio between the two, $|(w_p - \omega_p) / |\omega_p||$, is less than 1 (i.e., $-2 < w_p / |\omega_p| < 0$), background settling is considered dominant; otherwise, the interaction dynamics is considered complex. In particular, when $w_p / |\omega_p| \leq -2$, the complex flow-particle interaction dynamics causes the particle to settle much more quickly when compared with background settling (complex settling); while when $w_p / |\omega_p| \geq 0$, the complex flow-particle interaction dynamics causes the particles to rise rather than settle (rising). Figure 6 shows the spatial distribution in the *x-z* plane of $w_p / |\omega_p|$ at six selected times along with the particle velocity vectors. At the beginning, the particles in the upper fluid layers are forced to move backward to the left by the invasion of ambient fluid, while the particles in the lower part are driven forward by the flow (Figure 6(a)). At this time, particles close to the bottom boundary are mainly in the background settling regime ($-2 < w_p / |\omega_p| < 0$, shown as grey region), while most other particles are in the complex settling regime (shown as red region). Following the time evolution of the TC, the overall effect of fluid-particle interactions on the vertical grain motion becomes weaker and weaker, which is evident from the increasing proportion of the grey area in the core of the TC region (Figure 6). In addition, near the interface between the head and the ambient fluid, particles rise (shown as blue region). Comparing the particle velocity vectors in Figure 6 with the fluid velocity vectors in Figure 3, it can be seen that the two are highly correlated, that is, the two phases are in a coordinated motion, which is evidence of generally strong fluid-particle interactions.

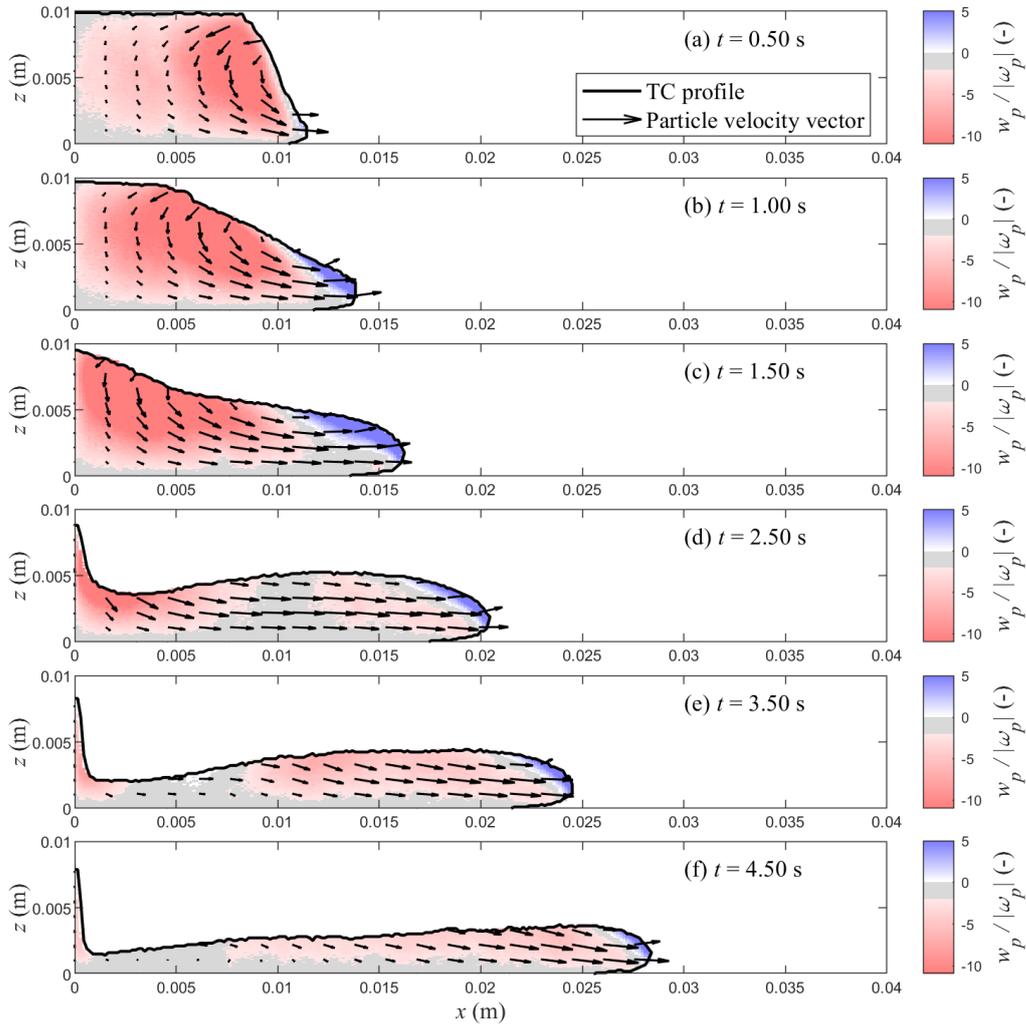

Figure 6. The spatial distributions in the *x-z* plane of the ratio of particle vertical velocity $w_p$ to absolute terminal settling velocity of a single particle $|\omega_p|$ (color maps) and the spatial distributions of particle velocity vectors (black arrows) in the *x-z* plane. For the color maps, red to white indicates the ratio $w_p / \omega_p$ from -11 to -2, grey indicates the ratio from -2 to 0, and white to blue indicates the ratio from 0 to 5. The solid line indicates the TC profile, defined by a particle concentration threshold

of $\alpha_p = 10^{-5}$.

Henceforth, particles that are located very close (within $1.5d_p$) to the bottom wall or deposited particles and nearly do not move ($|\mathbf{u}_p| < 10^{-4}$ m/s) are defined as deposited particles and the remaining particles as transported particles. (Note that this recursive definition allows for clusters of deposited particles in which only one particle of the cluster needs to be close to the bottom wall.) The spatial-temporal evolution of the average deposition thickness $h_{deposited}$ (volume of deposited particles per unit area) exhibits two distinct behaviors (Figure 7(a)): left of the gate ($x < 0.01$ m), there is little variation of $h_{deposited}$ beyond statistical noise, while right from the gate ($x > 0.01$ m), deposition at a given position $x$ begins only once the TC front has reached $x$, which is why, for a given time $t$, $h_{deposited}$ decreases with distance from the gate until it reaches zero (deep blue region). The total mass of deposited particles nondimensionalized by the total mass of particles $\widetilde{m}_{deposited}^{sum}$ gradually increases (Figure 7(b)), reaching close to 0.3 times the total number of particles at the end of the simulation duration, consistent with Necker et al. (2002). Interestingly, the deposition rate $D\widetilde{m}_{deposited}^{sum}/Dt$, when averaged over short-term noise, increases linearly with time throughout the simulation (Figure 7(c)).

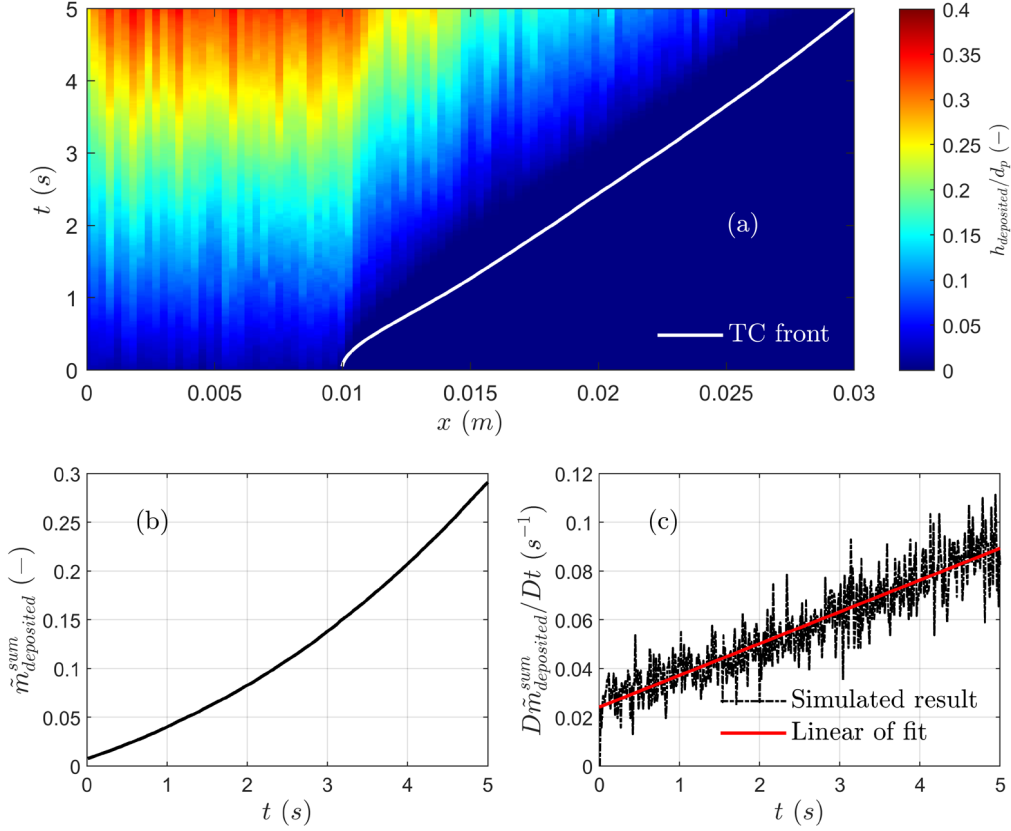

Figure 7. (a) Temporal and spatial distribution of average deposition thickness, (b) temporal variation of deposited particle mass, and (c) temporal variation of deposition rate.

**4 Fluid-particle interactions**

4.1 Characteristics of the average interaction forces and the implications for TC evolution

Figure 8 shows the temporal variation of the longitudinal components of the average forces (i.e., the drag force, the lift force, the added mass force, and the contact force) acting on transported particles (as defined in the previous section). Also included in Figure 8 are: (1) the sum of the average drag force, the lift force, and the added mass force, which is termed the "longitudinal interaction force", as well as (2) the sum of the interaction force and the contact force, which is termed the "total longitudinal force". The total longitudinal force can be used to judge whether particles are accelerated or decelerated by the flow on average. It can be seen that, for $t < 2.05$ s (Stage I), the total longitudinal force is positive, while for $t > 2.05$ s (Stage II), the total longitudinal force is negative. The key reason for this change is that the magnitude of the negative lift force becomes larger than the positive drag force at and after the stage transition. Furthermore, it can be seen that, throughout the whole process, the drag force is positive, indicating that fluid moves faster than particles ($u_f$ - $u_p$ > 0). The exception is the very beginning ($t$ < 0.30 s), where a short-term negative drag force results from the fact that, initially, many particles are located at high elevations (since they have not yet had enough time to settle down) where the longitudinal fluid velocity is negative (see Figure 3(a)). Opposite to the drag force, the lift force experiences a change from positive to negative. At the early phase of Stage I (0 < $t$ < 1.30 s), positive vorticity in the *x-z* plane (see Figure 4(a)) and positive vertical slip velocity (i.e., $w_f$ - $w_p$ > 0) make the lift force positive (see Eq. (11), the lift force is determined by both vorticity and slip velocity). At later stages (about $t$ > 1.30 s), the lift force decreases to zero and then becomes negative because more and more particles are affected by the negative vorticity near the bottom wall. Moreover, the added mass force exhibits a comparably large positive value in Stage I because the slip velocities $u_f$ - $u_p$ of most particles change strongly with time. However, once the rate of change slows down, the added mass force becomes negligible. During most of the process, the magnitude of the contact force remains substantially smaller than that of the other forces, so that the total force and the interaction force are roughly the same.

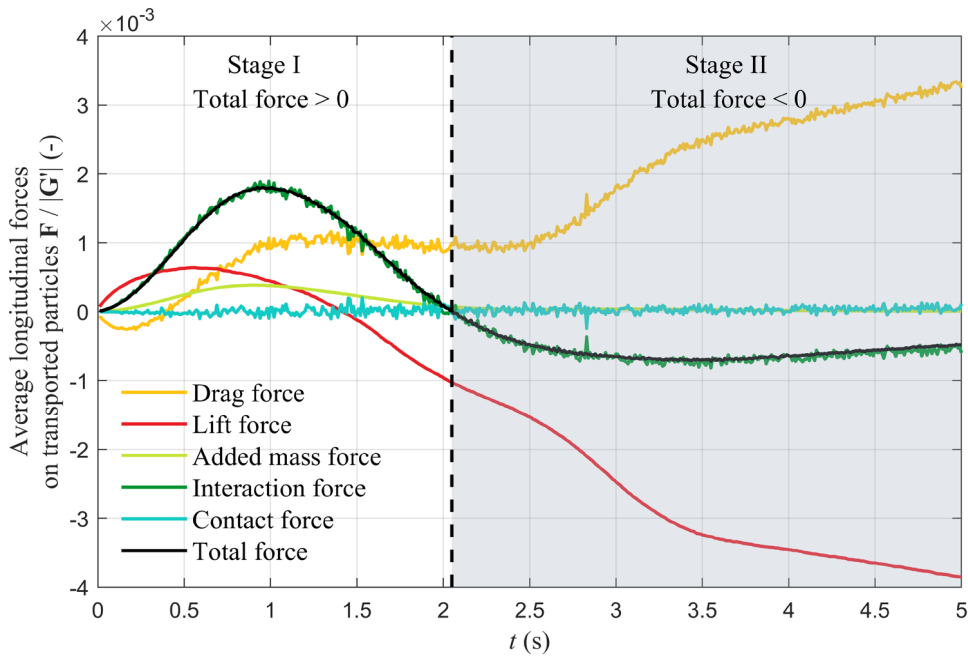

Figure 8. Time variation of the average longitudinal forces on transported particles. The sum of the longitudinal components of drag force, lift force, and added mass force represents the longitudinal interaction force. The sum of the longitudinal components of drag force, lift force, added mass force,

and contact force represents the total longitudinal force. $\mathbf{G}'$ is the effective gravity ($\mathbf{G}' = m\mathbf{g} - \frac{1}{6}\pi\rho_f d_p^3 \mathbf{g}$).

Figure 9 shows the temporal variations of the vertical components of the average forces acting on transported particles, including (a) the drag force, the lift force, the added mass force, the contact force, and the effective gravity force, which is the net force of the buoyancy and the gravity force; and (b) the total vertical force. It can be seen that the lift force, added mass force, and contact force are negligible when compared to the drag force (Figure 9(a)), which almost solely resists the effective gravity force and does not change significantly with time. The sum of both forces, and thus the total vertical force, initially increases and exceeds zero around $t = 0.91$ s (Figure 9(b)).

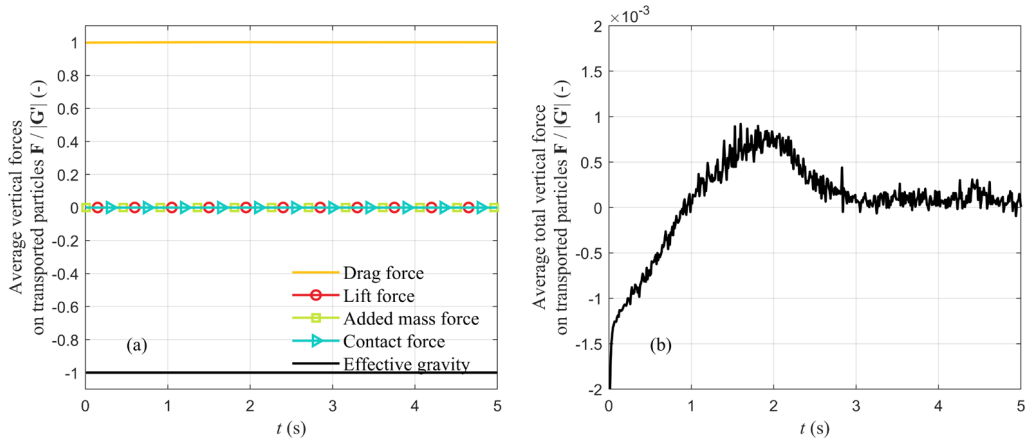

Figure 9. (a) Average vertical forces and (b) average total vertical force acting on transported particles as functions of time. The effective gravity in Figure 9(a) is the net force of the gravity and the buoyancy force. The total vertical force in Figure 9(b) is the net force of the drag force, lift force, added mass force, contact force, and effective gravity. $\mathbf{G}'$ is the effective gravity ($\mathbf{G}' = m\mathbf{g} - \frac{1}{6}\pi\rho_f d_p^3 \mathbf{g}$).

4.2 Head dynamics from the perspective of the interaction forces

The evolution of the TC is closely related to the head dynamics (Simpson, 1972; Nogueira et al., 2014). Here, the head length is set equal to 0.15 times the length of the TC (i.e., $0.15x_{front}$). We confirmed that other reasonable choices (e.g., $0.1x_{front}$ or $0.2x_{front}$) do not change the quality of the results (Appendix B). The average longitudinal forces on the transported head particles (including the drag force, lift force, added mass force, contact force, interaction force, and total force) are presented in Figure 10. According to whether total longitudinal force is positive or negative, the head evolution can also be divided into two stages. In Stage I (0~1.82 s), the total force on the head particles is positive, while in Stage II ($t > 1.82$ s), the total force is negative. Interestingly, the exact time differentiating the two stages is earlier for the head (1.82 s, see Figure 10) than that for the whole TC (2.05 s, see Figure 8). The reason is that the contact force, which plays no significant role for the average overall dynamics, plays a significant role for the average head dynamics due to the continuous impact of upstream particles on the TC head. In the initial stages, it is comparable to the other forces (sometimes even exceeding them).

The lift force on the head particles is always positive due to the positive vertical slip velocity (inferred by the positive vertical drag force on the head particles in Figure 11(a)) and the positive vorticity in the *x-z* plane (shown in Figure 4). This differs from the lift force in the whole TC, which is mostly negative for $t > 1.30$ s (Figure 8), because the expansion of the negative vorticity region has a weaker effect on the head area. While the average lift force in the whole region gradually drops to a negative value in Figure 8, the decreasing lift force in the head region does not change its direction and only gradually approaches zero at the end of the simulation duration.

Due to the large head lift force, the head particles move faster than the fluid, thus leading to a negative longitudinal slip velocity, so that the drag force on the head particles is always negative. These behaviors differ from those of the drag force and slip velocity in the whole TC region, which are positive on average. In addition, the average added mass force acting on the head transported particles exhibits a small positive value in 0~0.5 s, and a small negative value afterwards.

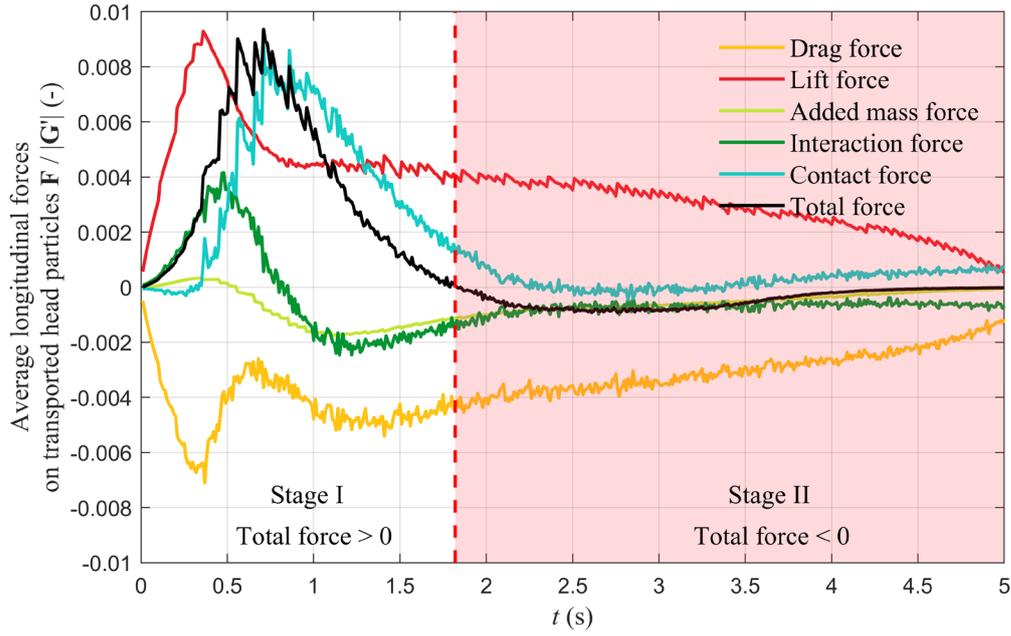

Figure 10. Temporal variation of the average longitudinal forces on transported head particles. The sum of the longitudinal components of drag force, lift force, and added mass force represents the longitudinal interaction force. The sum of the longitudinal components of drag force, lift force, added mass force, and contact force represents the total longitudinal force. $\mathbf{G}'$ is the effective gravity ( $\mathbf{G}' = m\mathbf{g} - \frac{1}{6}\pi \rho_f d_p^{\,3} \mathbf{g}$ ).

Figures 11(a) and 11(b) present the temporal evolution of the vertical components of the average forces acting on transported head particles. It can be seen that the key to preventing the TC head particles from falling down is the drag force (Figure 11(a)), like for TC particles in general (Figure 9(a)). The average total vertical force on transported head particles also appears to be negative first and then positive (Figure 11(b)). Nevertheless, the average total force on transported head particles earlier becomes positive (0.58 s in Figure 11(b)) compared with the whole TC (0.91 s in Figure 9(b)).

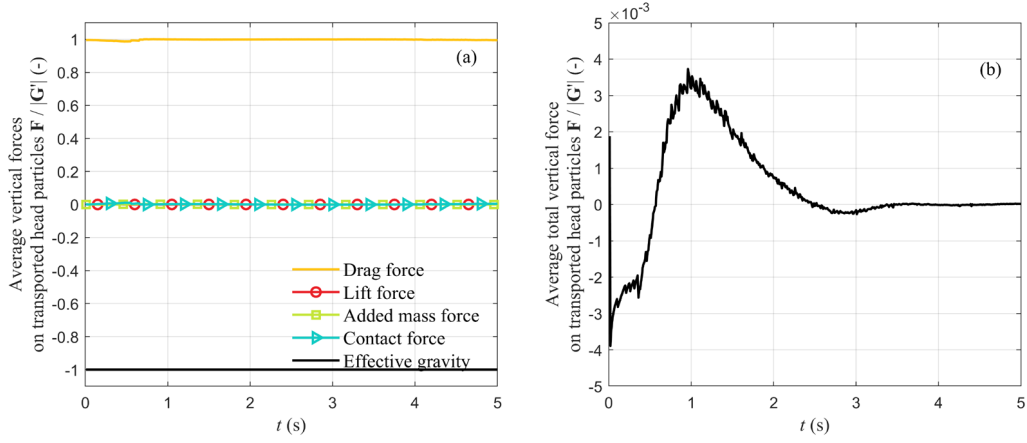

Figure 11. Temporal variations of (a) average vertical forces and (b) average total vertical force acting on transported head particles. The effective gravity in Figure 11(a) is the net force of the gravity and the buoyancy force. The total vertical force in Figure 11(b) is the net force of the drag force, lift force, added mass force, contact force, and effective gravity. $\mathbf{G}'$ is the effective gravity ($\mathbf{G}' = m\mathbf{g} - \frac{1}{6}\pi\rho_f d_p^3 \mathbf{g}$).

4.3 Process of energy conversion

Energy conversion processes are analyzed below. There are four energy constituents: the gravitational potential energy $E_p^p$ and the kinetic energy $E_k^p$ of the particle phase, and the potential energy $E_p^f$ and the kinetic energy $E_k^f$ of the fluid phase:

$$E_p^p(t) = \sum_{i=1}^{N_p} m_i g z_{p,i} \tag{24a}$$

$$E_k^p(t) = \sum_{i=1}^{N_p} \left( \frac{1}{2} m_i \left|\mathbf{u}_{p,i}\right|^2 + \frac{1}{2} I \left|\boldsymbol{\omega}_{p,i}\right|^2 \right) \tag{24b}$$

$$E_p^f(t) = \int_\Omega \alpha_f \rho_f g z dV \tag{24c}$$

$$E_k^f(t) = \int_\Omega \frac{1}{2} \alpha_f \rho_f \left|\mathbf{u}_f\right|^2 dV \tag{24d}$$

where $z_{p,i}$ is the elevation of the particle $i$ and $\Omega$ represents the whole simulation domain. The energy components in the system have the following relationship,

$$E_{Diss}(t) = E_{sum}(t_0) - E_{sum}(t) = E_p^p(t_0) - \left[E_p^p(t) + \Delta E_k^p(t) + \Delta E_p^f(t) + \Delta E_k^f(t)\right] \tag{25}$$

where $E_{Diss}$ is the dissipated energy, $E_{sum}$ is the total mechanical energy in the system, including that of the fluid phase and particle phase, $t_0$ is the initial time of the simulation, and $\Delta E_k^p$, $\Delta E_p^f$ and $\Delta E_k^f$ represent the change in particle kinetic energy ($\Delta E_k^p(t) = E_k^p(t) - E_k^p(t_0)$), in fluid potential energy ($\Delta E_p^f(t) = E_p^f(t) - E_p^f(t_0)$), and in fluid kinetic energy ($\Delta E_k^f(t) = E_k^f(t) - E_k^f(t_0)$), respectively. Figure 12 presents the temporal variation of these energy types non-dimensionalized by the particle gravitational potential energy at the initial time $E_p^p(t_0)$. It can be seen that the energy of the gravity currents mainly comes from the particle gravitational potential energy (Necker et al., 2002; Francisco et al., 2017), which decrease with time (Figure 12(a)). Two periods can be noted from Figure 12(a): before and after 2.0 s. Within 0~2.0 s, the particle gravitational potential energy decays by about 50% of its total (the rate of decrease is about 0.25 s$^{-1}$). A similar extent of reduction (decrease about 67%) in the particle gravitational potential energy has been reported by Necker et al. (2002). After 2.0 s, due to particle collapse (He et al., 2018), its decreasing rates slow down to 0.10 s$^{-1}$.

The particle gravitational potential energy is predominantly converted into fluid potential energy and dissipated energy (Figures 12(b)-12(d)). In contrast, only a very small fraction of the particle gravitational potential energy is converted into particle and fluid kinetic energy (Figure 12(c)). However, though very small, this energy conversion process, which occurs largely due to particle-fluid interactions, is what drives the complex TC dynamics, as shown in the previous subsections.

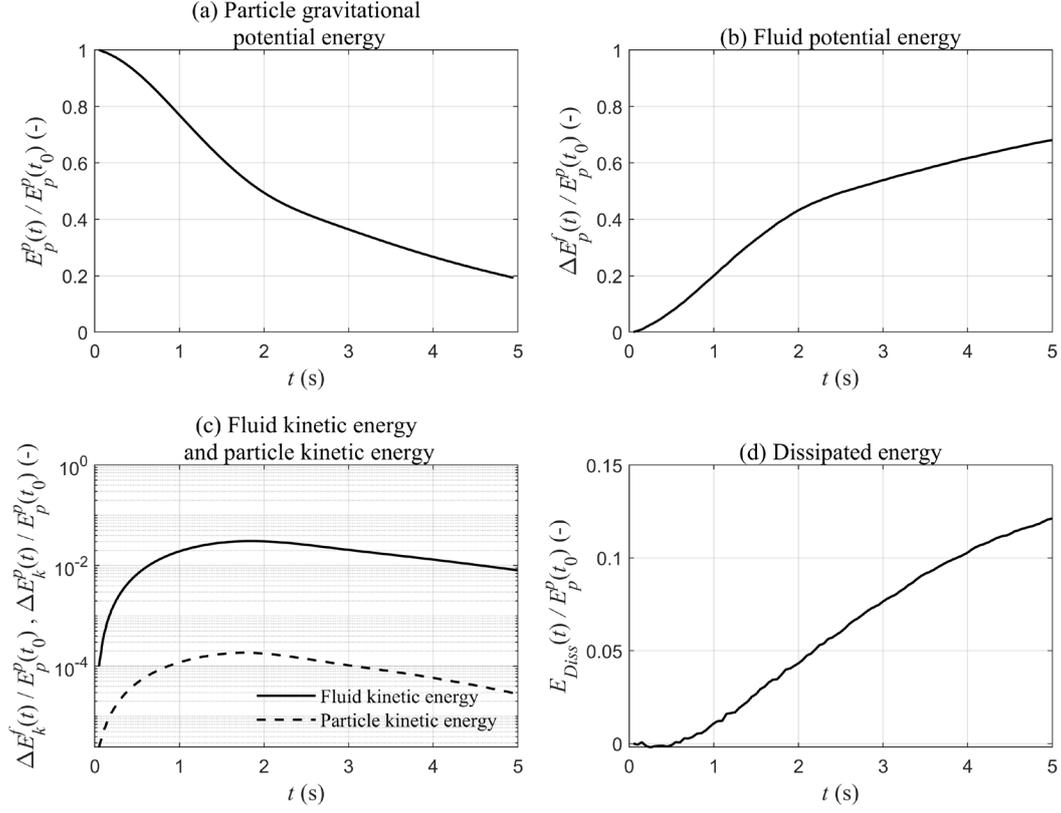

Figure 12. Temporal variations of different non-dimensionalized energy components.

## 5 Conclusions

In this paper, a LES-DEM model has been adopted to investigate the fluid-particle interactions in a lock-exchange turbidity current (TC) over a flat bed. The model predictions are consistent with measurements of the time evolution of the lock-exchange TC front position, fluid velocity profile, and particle concentration profile, and qualitatively consistent with previous findings regarding the fluid and particle flow. We have used the model to study the particle-fluid interactions during the TC evolution. In general, the evolution of the TC over a flat bed can be divided into two stages: a first stage in which particles are accelerating and a second stage in which they are decelerating. A more careful look reveals that the transition between these two stages occurs not at a fixed time but depends on the region of the TC one focuses on. For example, it occurs earlier for the head than for other parts of the TC. The two stages emerge due to temporal variations of the interaction forces between the particle and fluid phases. Among these, it is shown that the lift force, which is determined by both the vorticity field and the slip velocity between particles and fluids, plays a very important role. At the very beginning, longitudinal positive lift due to the strong positive vorticity promotes longitudinal particle transport. Afterwards, the longitudinal lift force decreases and eventually becomes negative, with a magnitude that even exceeds that of the positive longitudinal drag force, because more and more of the settling particles are affected by the negative vorticity near the bottom wall caused by surface friction. Interestingly, in spite of the complex behavior of the fluid-particle interaction forces and their role in TC evolution, only a very small fraction of the initial particle gravitational potential energy is actually transformed into TC kinetic energy (both particle and fluid). The effects of bed slope, sediment concentration as well as other key factors are reserved for future study.

**Appendix A. Influence of $Q$-criterion thresholds on coherent structures**

The spatial scale and local features of the coherent structures obtained with the $Q$-criterion are affected by the adopted $Q$ threshold. Figure A1 shows the coherent structures at 3.50 s using different $Q$ thresholds. It can be seen that the main coherent structures (large vortices) near the head extracted for different thresholds are similar and only differ in the spatial scale. The larger the threshold $Q$, the smaller the captured coherent structure. As for the fine fragmented coherent structures near the upper and lower interfaces of the TC, the number and spatial scale of these structures decrease to a certain extent with the increase of the threshold. Considering the purpose of capturing as many structures as possible while not overloading the images, we chose $Q = 0.5$ s$^{-2}$ as the threshold to capture the coherent structures in this study.

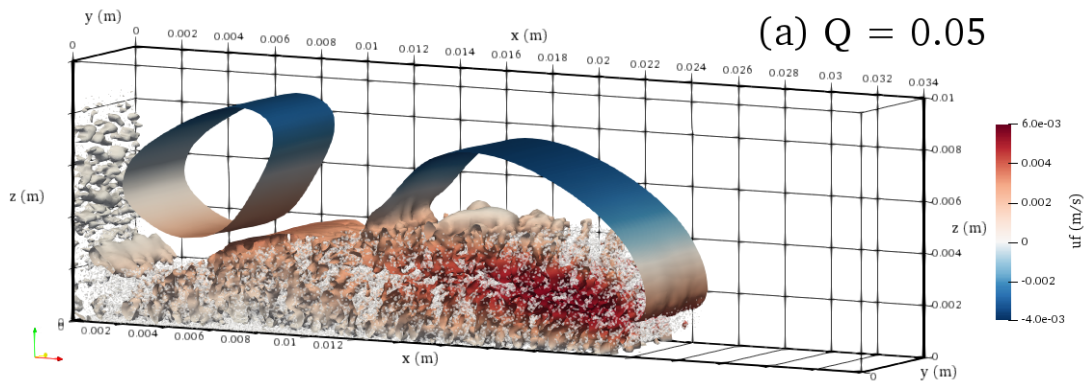

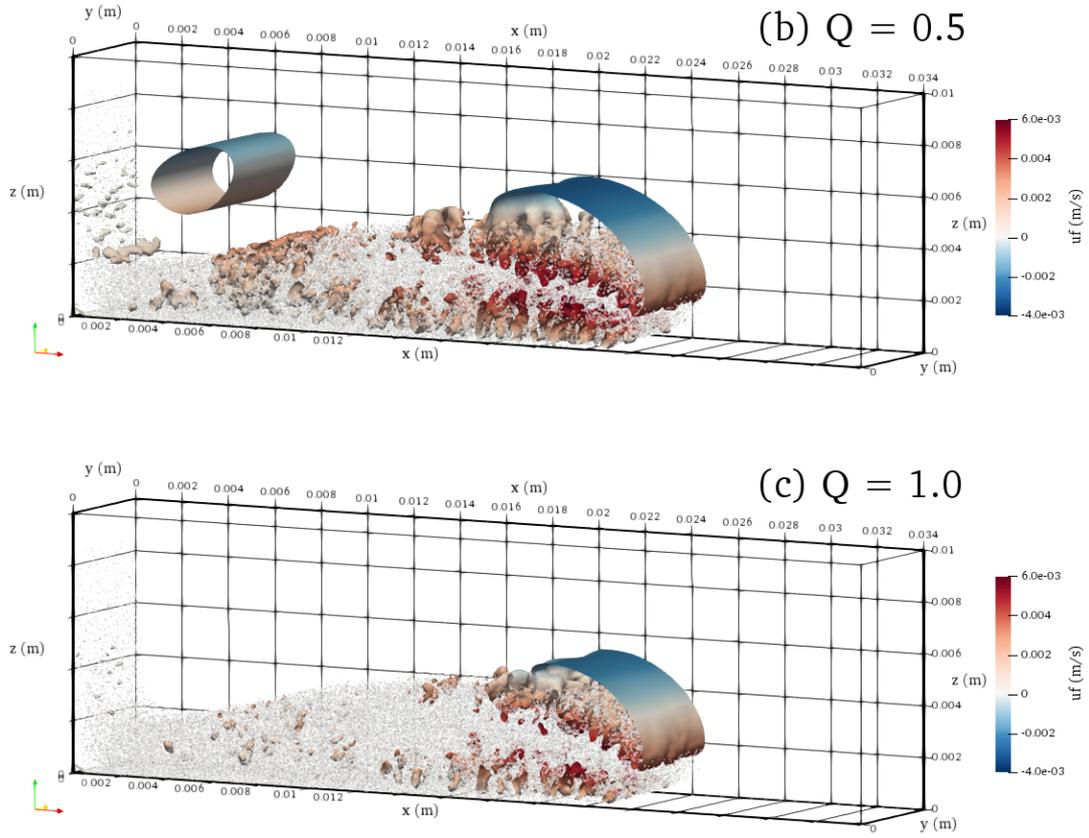

Figure A1. The coherent structures obtained using the $Q$-criterion when (a) $Q = 0.05$ s$^{-2}$, (b) $Q = 0.5$ s$^{-2}$, and (c) $Q = 1.0$ s$^{-2}$. The color of the iso-surface represents the fluid longitudinal velocity. The very small grey dots indicate particles.

**Appendix B. Sensitivity analysis of head length to average force on transported head particles**

Here, we discuss the effect of different head lengths on the average force acting on transported head particles. Three different head lengths are considered: $0.10 x_{front}$ (R1), $0.15 x_{front}$ (R2), and $0.20 x_{front}$ (R3), as shown in Figure B1. It can be seen that, for these different definitions, the time evolutions of the average force on the transported head particles are qualitatively very similar. That is, selecting different head lengths within this range ($0.10 \sim 0.20 x_{front}$) does not have an impact on the quality of the force analysis results.

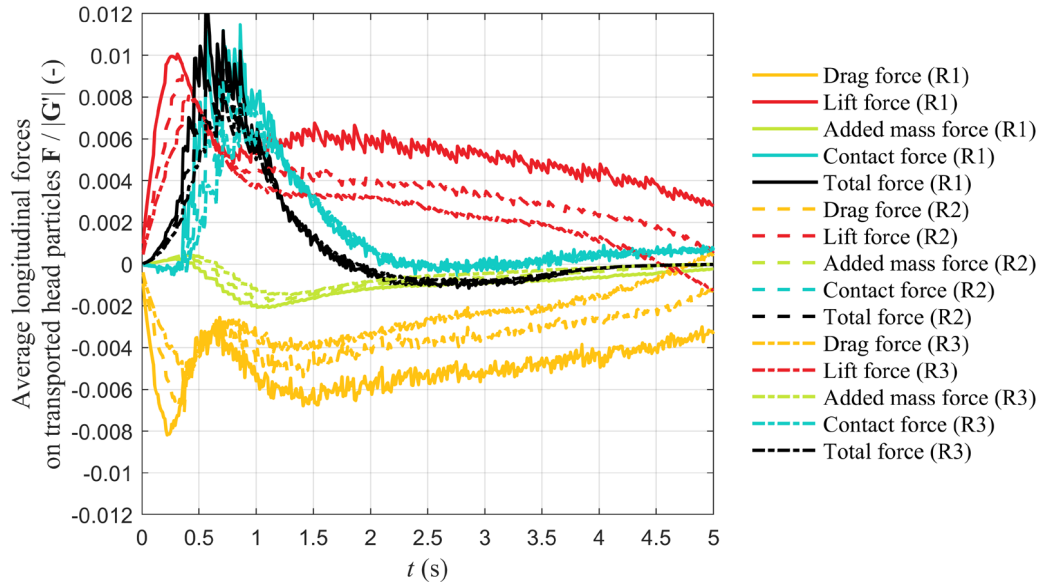

Figure B1. Temporal variation of the average longitudinal forces on transported head particles (drag force, lift force, added mass force, contact force and total force) under three head length definition methods (R1 for $0.10x_{front}$, R2 for $0.15x_{front}$, and R3 for $0.20x_{front}$).

**Acknowledgements**

This paper is supported by the National Key Research and Development Program of China (Grant NO. 2021YFF0501302), the National Natural Science Foundation of China (Nos. 12172331, 11872332), the Zhejiang Natural Science Foundation (No. LR19E090002), and the HPC Center of ZJU (ZHOUSHAN CAMPUS).